\documentclass{article}

\usepackage[cmex10]{amsmath}
\usepackage[pdftex]{graphicx}
\usepackage{subfigure}
\usepackage{amsthm}
\usepackage{amsfonts}
\def\C{{\mathbb C}}
\def\R{{\mathbb R}}

\def\N{{\mathbb N}}

\DeclareMathOperator{\Irr}{Irr}
\DeclareMathOperator{\tr}{tr}
\DeclareMathOperator{\Log}{Log}

\newtheorem{proposition}{Proposition}
\newtheorem{theorem}{Theorem}
\newtheorem{definition}{Definition}

\begin{document}

\title{Decompounding on compact Lie groups}
\author{Salem~Said$^{(1)}$, Christian~Lageman$^{(2)}$,\\
 Nicolas~Le~Bihan$^{(1)}$~
and~Jonathan~H.~Manton$^{(3)}$}
\date{(1): GIPSA-Lab / CNRS, Grenoble, France; \\
(2): Department of Electrical Engineering and Computer Science,\\
Universite de Liege, Belgium; \\
(3): Department of Electrical and Electronic Engineering,\\
The University of Melbourne, Australia.\\
\vspace{0.5cm}
{\small \tt Salem.said@gipsa-lab.grenoble-inp.fr\\
christian.lageman@montefiore.ulg.ac.be\\
nicolas.le-bihan@gipsa-lab.grenoble-inp.fr\\
jmanton@unimelb.edu.au}}
\maketitle

\begin{abstract}
Noncommutative harmonic analysis is used to solve a nonparametric estimation problem stated in terms of compound Poisson processes on compact Lie groups. This problem of \textit{decompounding} is a generalization of a similar classical problem. The proposed solution is based on a characteristic function method. The treated problem is important to recent models of the physical inverse problem of multiple scattering.
\end{abstract}

\section{Introduction}
This paper studies the following nonparametric
estimation problem.  Let $(X_n)_{n \geq 1}$ be \textit{i.i.d.}
$G$-valued random variables for some group $G$, and let $e$ denote
the identity element of $G$.  For example, $G$ might be the group
of $3 \times 3$ orthogonal matrices, in which case each $X_n$ would
be a random $3 \times 3$ orthogonal matrix and $e$ would be the
$3 \times 3$ identity matrix.  The process
$$
Y(t) = \prod^{N(t)}_{n=0} X_n, \qquad X_0 = e,
$$
where $N = (N(t))_{t \geq 0}$ is a Poisson process with parameter
$\lambda > 0$, is called a $G$-valued compound Poisson process.  If
$G$ is not commutative, the above products are taken to be ordered
from left to right, and $Y(t)$ is called a \textit{left} compound
Poisson process.  It is assumed that the random variables $X_n$ and
$N(t)$ are independent of each other, and for simplicity, it is
further assumed that the Poisson parameter $\lambda$ is known.  The
general problem is to estimate the distribution of the $X_n$ given
partial observations of one or more realisations of the compound
Poisson process $Y(t)$.  Of specific interest, is the case when
multiple realisations of $Y(T)$ are available, for some fixed time
instant $T > 0$.

The real numbers form a group, with addition being the group
operation.  Choosing $G$ to be this group results in the ordinary
compound Poisson process $y(t) = \sum_{n=0}^{N(t)} x_n$ where $x_0=0$
and $x_n$ for $n \geq 1$ are real-valued \textit{i.i.d.} random
variables. Estimating the distribution of the $x_n$ is known as
decompounding and has been well-studied~\cite{bg:03,egs:07}. In the present paper, decompounding
techniques are generalised to the case when $G$ is a noncommutative group. This
generalisation is non-trivial and requires ideas from noncommutative
harmonic analysis. Although group-valued compound Poisson processes
were introduced by Applebaum in~\cite{a:00}, the corresponding
decompounding problem has not been addressed in generality before.

This paper contributes to the relatively recent trend consisting in the application of noncommutative harmonic  analysis (\textit{i.e.} harmonic analysis on groups) to estimation and inverse problems. It addresses a nonparametric estimation problem stated in terms of compound Poisson processes on compact Lie groups. We refer to this as the problem of \textit{decompounding} on compact Lie groups, since it directly generalizes the classical problem of decompounding for scalar processes. This generalization is mathematically natural and is  motivated by the physical inverse problem of multiple scattering. In particular, this paper also contributes to the modelling of multiple scattering using compound Poisson processes.

Compound Poisson processes model the accumulation of rare events. As such, scalar compound Poisson processes are important tools in queuing and traffic problems and in risk theory. The classical problem of decompounding arises in the context of these processes. A functional approach to this problem is given by Buchman and Gr{\"u}bel~\cite{bg:03}. A characteristic function method is studied by Van Es \textit{et al.}~\cite{egs:07}. The applications of decompounding in queuing problems and risk theory are referenced in~\cite{bg:03}. We generalize this problem by considering decompounding on compact Lie groups. We approach this new problem by using noncommutative harmonic analysis to generalize the above mentioned method of~\cite{egs:07}. 

The important potential which noncommutative harmonic analysis holds for engineering problems is well illustrated in the book of Chirikjian and Kyatkin~\cite{ck}. Its importance to nonparametric estimation stems from the fact that it leads to the successful generalization of the highly important concept of characteristic function in probability. In mathematical research, this generalization was pioneered by Grenander~\cite{gre} and extensively developed by Heyer~\cite{hey}. It has received special attention in the engineering community. See Yazici~\cite{yazici:04} and the papers by Kim \textit{et
al.}~\cite{kk:08,kk:02,kr:01,kim:98}.

The paper is organized as follows. Section \ref{sec:hap} sets down the necessary background in harmonic analysis and characteristic functions on compact Lie groups. Section \ref{sec:cpp} introduces compound Poisson processes on compact Lie groups. In Section \ref{sec:decomp} we state the decompounding problem for these processes and present our approach based on noncommutative harmonic analysis. In Section \ref{sec:so(3)} we propose a model for multiple scattering based on compound Poisson processes on the rotation group $SO(3)$. Within this model, decompounding appears as a physical inverse problem. We apply our approach as described in Section \ref{sec:decomp} to this problem using numerical simulations. 

\section{Characteristic functions on compact Lie groups} \label{sec:hap}
Characteristic functions of scalar and vector-valued random variables are defined using the usual Fourier transform. Their extension to random variables with values on compact Lie groups owes to the tools of harmonic analysis on these groups. Our presentation of characteristic functions is adapted from~\cite{gre,liao}. Harmonic analysis on compact Lie groups is presented in more detail in recent papers~\cite{kk:08,yazici:04}. More thorough classical references thereon include~\cite{bd,dk}.

Let $G$ be a compact connected Lie group with identity $e$. We denote by $\mu$ the biinvariant normalized Haar measure on $G$.  Hilbert spaces of square integrable (with respect to $\mu$) complex and real-valued functions on $G$ are noted $L^2(G,\C)$ and $L^2(G,\R)$. A representation of $G$ is a continuous homomorphism $\pi\colon G \rightarrow GL(V)$ with $V$ a complex Hilbert space and $GL(V)$ the group of invertible bounded linear maps of $V$. It is called irreducible if any $G$-invariant subspace of $V$ is trivial i.e. equals $\{0\}$ or $V$. Two representations $\pi_i \colon G \rightarrow GL(V_i)$ --with $i = 1,2$-- are called equivalent if there exists an invertible bounded linear map $L:V_1 \rightarrow V_2$ such that  $L \circ \pi_1 = \pi_2 \circ L$. Using this relation, the set of irreducible representations of $G$ is partitioned into equivalence classes.

The central result of harmonic analysis on compact groups is the Peter-Weyl theorem. For the current context, it can be stated as follows. Let $\Irr(G)$ be the set of equivalence classes of irreducible representations of $G$. $\Irr(G)$ is a countable set. If $\delta \in \Irr(G)$ then we have the two following facts. All representations of the class $\delta$ have the same finite dimension $d_\delta$. There exists in this class a unitary representation $U^\delta$. Choosing one such representation we can suppose that 
$U^\delta \colon G \rightarrow SU(\C^{d_\delta})$ with $SU(\C^{d_\delta})$ the group of special unitary $d_\delta \times d_\delta$ matrices. We distinguish the unit representation $\delta_0 \in \Irr(G)$ where $U^{\delta_0}(g) = 1$ for all $g \in G$. With this choice being fixed, we can state the Peter-Weyl theorem.
\begin{theorem}[Peter-Weyl] \label{th:pw}
The functions $d_\delta^{1/2} U^\delta_{ij}$ taken for $\delta \in \Irr(G)$ and $i,j=1,\ldots,d_\delta$ form an orthonormal basis
of $L^2(G,\C)$.
\end{theorem}
Note that $U^\delta_{ij}$ is the usual notation for the matrix elements of $U^\delta$. For all $f \in L^2(G,\C)$ the theorem gives the Fourier pair
\begin{equation} \label{eq:fcoef}
 A_\delta = \int f(g)U^\delta(g)^\dagger d\mu(g)
\end{equation}
\begin{equation} \label{eq:fseries}
 f(g) = \sum_{\delta\in \Irr(G)} d_\delta \tr(A_\delta U^\delta(g)) 
\end{equation}
where $^\dagger$ denotes the Hermitian conjugate and $\tr$ the trace. The Fourier series (\ref{eq:fseries}) converges in $L^2(G,\C)$.

Consider the example $G = S^1$. It is possible to make the identification $\delta = 0, 1, \ldots$. Then $U^\delta(z) = z^\delta$ for  $z \in S^1$. Writting $z = e^{i \theta}$ for some $\theta \in [0,2\pi]$, this gives the classical Fourier expansion of periodic functions.

We consider random objects and in particular $G$-valued random variables defined on a suitable probability space $(\Omega,\mathcal{A},\mathbb{P})$. When referring to the probability density of such a random variable $X$, we mean a probability density $p_X \in L^2(G,\R)$ with respect to $\mu$. The characteristic function of a $G$-valued random variable is defined as follows. Compare to~\cite{gre}.
\begin{definition} \label{def:char}
Let $X$ be a $G$-valued random variable. The characteristic function of $X$ is the map $\phi_X$ given by
$$
\delta \mapsto \phi_X(\delta) = \mathbb{E}(U^\delta(X)) \;\;\;\delta\in\Irr(G)
$$
\end{definition}
Here $\mathbb{E}$ stands for expectation on the underlying probability space. For all $\delta \in \Irr(G)$, the expectation in the definition is finite since $U^\delta$ has unitary values. When $X$ has a probability density $p_X$ its characteristic function gives the Fourier coefficients of $p_X$ as in (\ref{eq:fcoef}). We have
$$
\phi_X(\delta) = \mathbb{E}(U^\delta(X)) = \int p(g)U^\delta(g)d\mu(g) \;\;\;\delta\in\Irr(G)
$$ 
The following proposition \ref{prop:facts} reminds the relation between characteristic functions and the concepts of convolution and convergence in distribution. It is a generalization of classical properties for scalar random variables. Remember that a sequence $(X_n)_{n \geq 1}$ of $G$-valued random variables is said to converge in distribution to a random variable $X$ if for all real-valued continuous function $f$ on $G$ we have
$$
\lim_n \mathbb{E}(f(X_n)) = \mathbb{E}(f(X))
$$
The proof of proposition \ref{prop:facts} is straightforward. See~\cite{gre}.
\begin{proposition}\label{prop:facts}
The following two properties hold.
\begin{enumerate}
\item \label{facts:a} Let $X$ and $Y$ be independent $G$-valued random variables and let $Z = XY$. We have for all $\delta \in \Irr(G)$
$$
\phi_{Z}(\delta) = \phi_{X}(\delta) \phi_{Y}(\delta)
$$
\item \label{facts:b} A sequence $(X_n)_{n \geq 1}$ of $G$-valued random variables converges in distribution to a random variable $X$ \textit{iff} for all $\delta \in \Irr(G)$
$$
\lim_n \phi_{X_n}(\delta) = \phi_X(\delta)
$$
\end{enumerate}
\end{proposition} 
In order to solve our estimation problem in section \ref{sec:decomp} we will require random variables to have certain symmetry properties. We deal with these properties here. The following analysis draws on Liao~\cite{liao,liao:04}. 

We will say that a $G$-valued random variable $X$ is inverse invariant if $X \stackrel{d}{=} X^{-1}$. We will say that it is conjugate invariant if for all $k \in G$ we have that $X \stackrel{d}{=} kXk^{-1}$. As usual $\stackrel{d}{=}$ denotes equality in distribution. The following proposition \ref{prop:sym} characterizes these two symmetry properties in terms of characteristic functions. It will be important to remember that for any two $G$-valued random variables $X$ and $Y$ we have $X \stackrel{d}{=} Y$ \textit{iff} $\phi_X = \phi_Y$. This results from the completeness of the basis given by the $U^\delta$ as stated in the Peter-Weyl theorem~\cite{gre}.
\begin{proposition} \label{prop:sym}
The following properties hold.
\begin{enumerate}
\item \label{sym:a} $X$ is inverse invariant \textit{iff} for all $\delta \in \Irr(G)$ we have that $\phi_X(\delta)$ is Hermitian.
\item \label{sym:b} Let $X$ be inverse invariant. If $X_1, \ldots, X_n$ are independent copies of $X$ then the product $X_1 \ldots X_n$ is inverse invariant.
\item \label{sym:c}  $X$ is conjugate invariant \textit{iff} for all $\delta \in \Irr(G)$ we have that $\phi_X(\delta) = a_\delta I_{d_\delta}$ where $a_\delta\in\C$ and $I_{d_\delta}$ is the ${d_\delta} \times {d_\delta}$ identity matrix. 
\item \label{sym:e} If $X$ and $Y$ are independent and conjugate invariant then $XY$ is conjugate invariant. 
\item \label{sym:d} $X$ is conjugate invariant \textit{iff} for all $G$-valued random variable $Y$ independent of $X$ we have
$XY \stackrel{d}{=} YX$.
\end{enumerate}
\end{proposition}
\begin{proof}
\begin{enumerate}
\item Note that for all $\delta \in \Irr(G)$ we have by the homomorphism property of $U^\delta$ and the fact that it has
unitary values
$$
\phi_{X^{-1}}(\delta) = \mathbb{E}(U^\delta(X^{-1})) = \mathbb{E}(U^\delta(X))^\dagger = \phi_X(\delta)^\dagger
$$
\item This follows from \ref{sym:a} of proposition \ref{prop:sym} and \ref{facts:a} of proposition \ref{prop:facts}, since the powers of a Hermitian matrix are Hermitian.
\item Note that for all $k \in G$ we have that $X \stackrel{d}{=} kXk^{-1}$ \textit{iff} for all $\delta \in \Irr(G)$
$$
\mathbb{E}(U^\delta(X)) = \mathbb{E}(U^\delta(kXk^{-1})) = U^\delta(k)\mathbb{E}(U^\delta(X))U^\delta(k)^\dagger
$$
identifying $\phi_X$ on both sides, this becomes
$$
\phi_X(\delta) = U^\delta(k)\phi_X(\delta)U^\delta(k)^\dagger
$$
If this relation is verified for all $k \in G$ then $\phi_X(\delta)$ is a multiple of $I_{d_\delta}$. This follows by Schur's lemma~\cite{bd}. 
\item This follows from \ref{sym:c} of proposition \ref{prop:sym} and \ref{facts:a} of proposition \ref{prop:facts}.
\item The \textit{if} part follows by setting $Y = k \in G$ for arbitrary $k$. The \textit{only if} part follows from \ref{sym:c} of proposition \ref{prop:sym} and \ref{facts:a} of proposition \ref{prop:facts}. 
 \end{enumerate}
\end{proof}
\ref{sym:a} of proposition \ref{prop:sym} motivates a practical recipe for generating inverse invariant random variables from general random variables. Let $X$ and $Y$ be $G$-valued random variables. Suppose $X$ and $Y$ are independent with $Y \stackrel{d}{=} X^{-1}$. It can be verified by \ref{sym:a} of proposition \ref{prop:sym} that $XY \stackrel{d}{=} YX$ and that both these products are inverse invariant. In practice, if we have generated $X$ then we can immediately generate $Y$ as above. In this way an inverse invariant $XY$ or $YX$ is generated from $X$.

\section{Compound Poisson Processes} \label{sec:cpp}
Compound Poisson processes on groups naturally generalize scalar compound Poisson processes. They are introduced by Applebaum in~\cite{a:00}. Let us start by reminding the definition of scalar compound Poisson processes. Let $N = (N(t))_{t \geq 0}$ be a Poisson process with parameter $\lambda > 0$. Suppose $(x_n)_{n \geq 1}$ are \textit{i.i.d.} $\R$-valued random variables. Suppose the family $(x_n)_{n \geq 1}$ is itself independent of $N$. The following process $y$ is said to be a compound Poisson process
$$
y(t) = \sum^{N(t)}_{n=0} x_n
$$
$G$-valued compound Poisson processes are defined by analogy to this formula. We continue with the process $N$. Let $(X_n)_{n \geq 1}$ be \textit{i.i.d.} $G$-valued random variables and suppose as before that the family $(X_n)_{n \geq 1}$ is independent of $N$. The following process $Y$ is said to be a $G$-valued left compound Poisson process
$$
Y(t) = \prod^{N(t)}_{n=0} X_n
$$
We understand that products are ordered from left to right. It is possible to obtain a right compound Poisson process by considering
$Y(t)^{-1}$ instead. Thus the two concepts are equivalent. See~\cite{liao,a:00}. 

Before going on, we make the following remark on the above definition of compound Poisson processes. This definition was stated for $G$ a compact connected Lie group. This topological and manifold structure of $G$ is not necessary for the definition, which can be stated in its above form for any group with a measurable space structure. The compact connected group structure of $G$ allows us to use the Peter-Weyl theorem and characteristic functions. The Lie group structure allows the introduction of Brownian noise in Section \ref{sec:decomp}.   

We wish to summarize the symmetry properties of the random variables $Y(t)$ for $t \geq 0$. Note first that for all $t \geq 0$, $Y(t)$ does not have a probability density. Indeed, for all $t \geq 0$ we have $\mathbb{P}(Y(t) = e) \geq \mathbb{P}(N(t) = 0) =
e^{-\lambda t}$. It follows that $Y(t)$ has an atom at $e$. In the absence of a probability density, we study $Y(t)$ for $t \geq 0$ using its characteristic function. This is given in the following Proposition \ref{prop:char} which can be seen to immediately generalize the well known formula for scalar compound Poisson processes. This proposition follows~\cite{liao,a:00}. 
\begin{proposition} \label{prop:char}
For all $t \geq 0$ the characteristic function $\phi_{Y(t)}$ of $Y(t)$ is given by
\begin{equation}\label{eq:char}
\phi_{Y(t)} (\delta) = \exp(\lambda t (\phi_X(\delta) - I_{d_\delta}))
\end{equation}
for $\delta \in \Irr(G)$, where $\phi_X \equiv \phi_{X_1}$. 
\end{proposition}
\begin{proof} Let $t \geq 0$. $\phi_{Y(t)}$ can be calculated by conditioning over the values of $N(t)$. Using the independence of $N$ and $(X_n)_{n \geq 1}$ we have for $\delta \in \Irr(G)$
$$
\phi_{Y(t)} (\delta) = e^{- \lambda t} \sum_{n \geq 0} \frac{(\lambda t)^n}{n!} \mathbb{E}\prod^n_{m=0} U^\delta(X_m) 
$$
Using the fact that $(X_n)_{n \geq 1}$ are \textit{i.i.d.} it is possible to replace
$$
\mathbb{E} \prod^n_{m=0}U^\delta(X_m) = \prod^n_{m=0} \mathbb{E}(U^\delta(X_m)) = \phi_X(\delta)^n
$$
the proposition follows by rearranging the sum.
\end{proof}
Combining Propositions \ref{prop:char} and \ref{prop:sym} we have the following proposition. It states that for all $t \geq 0$ the symmetry properties of $Y(t)$ are the same as those of the $X_n$.
\begin{proposition} \label{prop:cppsym}
For all $t \geq 0$ we have
\begin{enumerate}
\item If $X_1$ is inverse invariant then so is $Y(t)$.
\item If $X_1$ is conjugate invariant then so is $Y(t)$.
\end{enumerate}
\end{proposition}
We end this section with Proposition \ref{prop:uniformize}. It gives a property of uniformization of the distribution of $Y(t)$ as $t \uparrow \infty$. This is similar to the behavior of the products $X_1 \ldots X_n$ for $n \uparrow \infty$, see \cite{gre}. For a more general version of Proposition \ref{prop:uniformize} see~\cite{liao,liao:04}. We say that a $G$-valued random variable $X$ is supported by a measurable subset $S$ of $G$ if $\mathbb{P}(X \in S) = 1$. If $X$ and $X^\prime$ are $G$-valued random variables with $X \stackrel{d}{=} X^\prime$ then $X$ is supported by $S$ \textit{iff} $X^\prime$ is supported by $S$. In Proposition \ref{prop:uniformize}, $U$ is a $G$-valued random variable with probability density identically equal to $1$. That is, $U$ is uniformly distributed on $G$.  
\begin{proposition} \label{prop:uniformize}
If $X_1$ is not supported by any closed proper subgroup $S$ of $G$ or coset $gS$, $g\in G$ of such a subgroup then $Y(t)$ converges in distribution to $U$ as $t \uparrow \infty$.
\end{proposition}
\begin{proof} Under the conditions of the proposition we have for all for all $\delta \neq \delta_0$ that the eigenvalues of $\phi_X(\delta)$ are all $< 1$ in modulus~\cite{gre}. It follows that the eigenvalues of $\phi_X(\delta)-I_{d_\delta}$ all have negative real parts. Thus when $\delta \neq \delta_0$ we have by (\ref{eq:char}) that $\phi_{Y(t)}(\delta) \rightarrow 0$ as $t \uparrow \infty$. Moreover, it is immediate that $\phi_{Y(t)}(\delta_0) = 1$ for  $t \geq 0$. We conclude using \ref{facts:b} of Proposition \ref{prop:facts}. Note that~\cite{bd}
$$
\phi_U(\delta) = \int U^\delta(g)d\mu(g) = 0 \;\;\;\delta \neq \delta_0
$$ 
and $\phi_U(\delta_0) = 1$ trivially.
\end{proof}

\section{Decompounding} \label{sec:decomp}
In existing literature, \textit{decompounding} refers to a set of nonparametric estimation problems involving scalar compound Poisson processes~\cite{bg:03,egs:07}. In this section we will consider the generalization of these problems to compound Poisson processes on compact Lie groups. The new problems can be stated in the notation of Section \ref{sec:cpp}. We refer to them also as decompounding problems. As in the scalar case, they consist in estimation of the common probability density (supposed to exist) of the random variables $X_n$ from observations of the process $Y$. The unknown common probability density of the $X_n$ will be noted $p$. We are unaware of any work on similar problems for vector-valued compound Poisson processes. Our consideration of compact Lie groups is motivated by the applications presented in Section \ref{sec:so(3)}.

\subsection{Typology of decompounding problems} \label{subsec:typology}
Several decompounding problems can be stated, depending on the nature of the observations made of $Y$~\cite{egs:07}. Decompounding is performed from \textit{high frequency} observations if an individual trajectory of the process $Y$ is observed over time intervals $[0,T]$ where $T \uparrow \infty$. It is performed from \textit{low frequency} observations if \textit{i.i.d.} observations are made of the random variable $Y(T)$ for a fixed $T \geq 0$. 

Decompounding from high and low frequency observations lead to different difficulties. For high frequency observations, the problem is greatly simplified if the assumption is made that $X_n$ does not take the value $e$, for any $n \geq 1$. With probability $1$, a trajectory of $N$ has infinitely many jumps over $t \geq 0$. Under the assumption we have made, all these jumps correspond to jumps of $Y$ which we do observe. The jumps of $Y$ then give \textit{i.i.d.} observations of $X_1$ and the average time between these jumps is $1/\lambda$. In particular, it is important for high frequency observations to take the limit $T \uparrow \infty$.

Low frequency observations do not give direct access to $\lambda$. In scalar decompounding from low frequency observations, $\lambda$ is often assumed to be known~\cite{bg:03,egs:07}. In the context of a compact group $G$, Proposition \ref{prop:uniformize} leads to a difficulty that does not appear in scalar decompounding. Under the conditions of this proposition, if low frequency observations are made at a sufficiently large time $T$ then these observations will be uniformly distributed on $G$ and will have no memory of the random variables $X_n$.  

A third intermediate type of observations is possible. It is possible to make observations of an individual trajectory of $Y$ at regular time intervals $T, 2T, \ldots$. This is in fact equivalent to low frequency distributions. Remember that $N$ is a L{\'e}vy process, \textit{i.e.} has independent stationary increments. Moreover we have that the $(X_n)_{n \geq 1}$ are \textit{i.i.d}. Using this, it is possible to prove that the $G$-valued random variables 
$$
Y(T), Y(T)^{-1}Y(2T), Y(2T)^{-1}Y(3T)\ldots
$$
are \textit{i.i.d}. Thus our observations are \textit{i.i.d.} observations of $Y(T)$. This remark refers to the fact that $Y$ is a left L{\'e}vy process in $G$~\cite{liao}. We do not develop this here. 

\subsection{Noise model for low frequency observations} \label{subsec:noisemodel}
We will consider decompounding from low frequency observations. $T \geq 0$ is fixed and \textit{i.i.d.} observations $(Z_n)_{n \geq 1}$ of a noisy version $Z$ of $Y(T)$ are available. $Z$ is given by $Y$ corrupted by multiplicative noise. We have the noise model
\begin{equation} \label{eq:noisemodel}
Z = M Y(T) 
\end{equation}
where $M$ is independent of $Y$. By \ref{facts:a} of Proposition \ref{prop:facts} we have for the characteristic function of $Z$
$$
\phi_Z = \phi_M \phi_{Y(T)}
$$
The noise model is equivalent to having an initial value $Y(0) = M$ with a general distribution. We consider the case of Brownian noise. The characteristic function of $M$ is then given by~\cite{liao,kk:08} 
$$
\phi_M(\delta) = \exp\left(-\lambda_\delta \frac{\sigma^2}{2}\right) I_{d_\delta}
$$
where $\sigma^2$ is a variance parameter and for $\delta \in \Irr(G)$ the constant $\lambda_\delta$ is the corresponding eigenvalue of the Laplace-Beltrami operator. In particular, $\lambda_{\delta_0} = 0$ and $\lambda_\delta > 0$ for $\delta \neq \delta_0$. It is clear from \ref{sym:c} of Proposition \ref{prop:sym} that $M$ is conjugate invariant. It follows by \ref{sym:e} of Proposition \ref{prop:sym} that, as far as the distribution of $Z$ is concerned, left and right multiplication of $Y(T)$ by the noise $M$ are indifferent. 

It is possible to construct a $G$-valued process $\zeta$ such that $Z \stackrel{d}{=} \zeta(T)$. The corresponding construction is well known in the theory of group-valued L{\'e}vy processes and is referred to as interlacing~\cite{a:00,liao}. Here we only state this construction. Let $W$ be a Brownian motion on $G$ independent of $N$ and with variance parameter $\bar{\sigma}^2$. This is a process with continuous paths and independent stationary increments. Moreover, $W(0) = e$ and for $\delta \in \Irr(G)$
$$
\phi_{W(t)}(\delta) = \exp\left(-\lambda_\delta \frac{\bar{\sigma}^2}{2}t \right) I_{d_\delta}
$$
Let $T_0 = 0$ and suppose $(T_n)_{n \geq 1}$ are the jump times of $N$. The interlaced process $\zeta$ is defined as follows. We have $\zeta(0) = e$. For $t > 0$ and $n \geq 1$ we have
$$
 \zeta(t) = \zeta(T_{n-1})W(T_{n-1})^{-1}W(t) \hspace{0.35cm} \text{ on } \lbrace T_{n-1} \leq t < T_n \rbrace 
$$
where the following formula holds at each time $T_n$ (here $\zeta(T_n-)$ denotes the left limit at $T_n$)
$$
 \zeta(T_n)  =  \zeta(T_n-)X_n 
$$
This definition is sufficient, since $T_n \uparrow \infty$ almost surely. The term interlacing comes from the fact that the trajectories of $\zeta$ are obtained by introducing the jumps of $Y$ into the trajectories of $W$ as these jumps occur. The trajectories of $W$ are thus interlaced with the jumps of $Y$. 

For $t \geq 0$ the characteristic function of $\zeta(t)$ is given by
\begin{equation} \label{eq:charnoise0}
\phi_{\zeta(t)}(\delta)= 
\exp\left( t\lambda\phi_X(\delta)  - tI_{d_\delta} \left( \lambda + \frac{\lambda_\delta\bar{\sigma}^2}{2} \right) \right)
\end{equation}
for $\delta \in \Irr(G)$. It follows that we have $Z \stackrel{d}{=} \zeta(T)$ if $T\bar{\sigma}^2 = \sigma^2$. 

Although we do not deal with the case of high frequency observations we would like to end this subsection with a remark on the role of noise in this case. The trajectories of the interlaced process $\zeta$ are noisy versions of the trajectories of $Y$. However, these trajectories have the same jumps as the trajectories of $Y$. In this sense, high frequency observations are unaltered by noise.

\subsection{A characteristic function method} \label{subsec:estimators}
We present a characteristic function method for decompounding from low frequency observations. This method extends a similar one considered in~\cite{egs:07}. In carrying out this extension, we are guided by the properties of characteristic functions on $G$ presented in Section \ref{sec:hap}. Our observations $(Z_n)_{n \geq 1}$ and noise model (\ref{eq:noisemodel}) were described in \ref{subsec:noisemodel}. We aim to estimate the common density $p$ of the $X_n$. A characteristic function method consists in constructing nonparametric estimates for $p$ from parametric estimates for its Fourier coefficients $\phi_X(\delta)$ given for $\delta \in \Irr(G)$. See~\cite{kk:08}. 

We suppose that $\lambda$ and $\sigma^2$ are known. Equation (\ref{eq:charnoise0}) can be copied as follows
\begin{equation} \label{eq:charnoise}
\phi_{Z}(\delta)= 
\exp\left( T \lambda\phi_X(\delta)  - T \bar{\lambda} I_{d_\delta}  \right) \;\;\;\delta\in\Irr(G)
\end{equation}
where $\bar{\lambda}$ is a constant determined by $\lambda$ and $\sigma^2$. We refer to this transformation $\phi_X \mapsto \phi_Z$ as the compounding transformation. Decompounding will involve local inversion of the compounding transformation. This is clearly related to inversion of the matrix exponential in a neighborhood of $\phi_Z(\delta)$ for all $\delta \in \Irr(G)$. Rather than deal with this problem in general, we make the following simplifying hypothesis. 
\begin{center}
\textbf{Hypothesis:} $X_1$ is inverse invariant.
\end{center}
For all $\delta \in \Irr(G)$ we have by applying \ref{sym:a} of Proposition \ref{prop:sym} and (\ref{eq:charnoise}) to this hypothesis that $\phi_Z(\delta)$ is Hermitian positive definite. Note $\Log$ the unique Hermitian matrix logarithm of a hermitian positive definite matrix. We can now express the inverse of the compounding transformation. From equation (\ref{eq:charnoise}) it follows that
\begin{equation} \label{eq:decompounding1}
 \phi_X(\delta) = \frac{1}{T\lambda} \Log\left[\phi_{Z}(\delta)\right] + 
                  \left(\bar{\lambda}/\lambda\right) I_{d_\delta} \;\;\;\delta\in\Irr(G)
\end{equation}
Let $\delta \in \Irr(G)$. It follows from definition \ref{def:char} that empirical estimates of $\phi_Z(\delta)$ based on the observations $(Z_n)_{n \geq 1}$ are unbiased and consistent. This is a simple consequence of the strong law of large numbers. See for example~\cite{Ka:02}. In order to estimate $\phi_X(\delta)$ using (\ref{eq:decompounding1}) it is then important to ensure that the empirical estimates of $\phi_Z(\delta)$ are asymptotically Hermitian positive definite.  

We start by defining the empirical estimates $\hat{\phi}^n_Z(\delta)$ for $\delta \in \Irr(G)$ and $n \geq 1$
$$
\hat{\phi}^n_Z(\delta) = \frac{1}{2n} \sum^n_{m=1} \left( U^\delta(Z_m) + U^\delta(Z_m)^\dagger \right)
$$
Hermitian symmetrization of empirical estimates is necessary for the application of (\ref{eq:decompounding1}). Since it is a projection operator, this symmetrization moreover contributes to a faster convergence of the $\hat{\phi}^n_Z(\delta)$ to $\phi_Z(\delta)$. 

Continuous dependence of the spectrum of a matrix on its coefficients is a classical result in matrix analysis. Several more or less sophisticated versions of this result exist~\cite{vlg:89}. For a remarkably straightforward statement see~\cite{us:77}. For a complex matrix $C$ we will note $\lambda(C)$ its spectrum. For each $\delta \in \Irr(G)$ and $n \geq 1$ define the event $R^n_\delta$ by
$$
R^n_\delta =\lbrace \lambda(\hat{\phi}^n_Z(\delta)) \subset ]0,\infty[ \rbrace
$$
For $\delta \in \Irr(G)$, the sequence $(R^n_\delta)_{n \geq 1}$ controls the convergence of the spectra of the empirical estimates $\hat{\phi}^n_Z(\delta)$. In particular,
$$
\mathbb{P}(\cup_{n \geq 0} \cap_{m \geq n} R^m_\delta) = \lim_n \mathbb{P}(\cap_{m \geq n} R^m_\delta) = 1
$$ 
Using the events $R^n_\delta$ we can write down well defined estimates of $\phi_X$. These are noted $\hat{\phi}^n_X(\delta)$ for $\delta \in \Irr(G)$ and $n \geq 1$
$$
\begin{array}{rclrl}
\hat{\phi}^n_X(\delta) & = & 0 &  \text{on} &  \Omega - R^n_\delta \\
\hat{\phi}^n_X(\delta) & = & \frac{1}{T\lambda} \Log\left[\hat{\phi}^n_Z(\delta)\right] + 
                             \left(\bar{\lambda}/\lambda\right) I_{d_\delta} & \text{on} &    R^n_\delta 
\end{array}
$$
This expression gives our parametric estimates for the Fourier coefficients of $p$. We use them to construct nonparametric estimates based on an expression of the form (\ref{eq:fseries}). Let $(\Gamma_l)_{l \geq 1}$ be an increasing sequence of finite subsets $\Gamma_l \subset \Irr(G)$ with the limit $\cup_{l \geq 1} \Gamma_l = \Irr(G) - \lbrace \delta_0 \rbrace$. Let $K \geq 0$ and for each $\delta \in \Irr(G)$ note 
$$
f_\delta = d_\delta e^{-K\lambda_\delta}
$$ 
For $n \geq 1$ and $l \geq 1$ our nonparametric estimate $\hat{p}^n_l$ is given by
\begin{equation} \label{eq:nonparam}
\hat{p}^n_l(g) = 1 + \sum_{\delta \in \Gamma_l} f_\delta \tr\left( \hat{\phi}^n_X(\delta) U^\delta(g)^\dagger \right) \;\;\;g \in G
\end{equation}
The subscript $l \geq 1$ corresponds to a cutoff or smoothing parameter. Indeed, infinitely many representations are excluded from the sum over $\Gamma_l$. A more complete expression of this fact appears in~\cite{kk:08}. When $K > 0$ the coefficients $f_\delta$ form a convolution mask ensuring that the estimates $\hat{p}^n_l$ can be taken to converge to a smooth probability density. We make this more precise in \ref{subsec:conv}.  

It is usual to rewrite expressions similar to (\ref{eq:nonparam}) in terms of a group invariant kernel. See~\cite{kk:08,kk:02}. Such a transformation is not possible here due to the indirect nature of our observations. This is in particular related to the more involved form of the $\hat{\phi}^n_X(\delta)$ as given above.

\subsection{Convergence of parametric and nonparametric estimates} \label{subsec:conv}
Here we discuss the convergence of the parametric and nonparametric estimates given in \ref{subsec:estimators}. Our argument is presented in the form of Propositions \ref{prop:convparam} and \ref{prop:convnonparam} below. Proposition \ref{prop:convparam} gives the consistency of the parametric estimates $\hat{\phi}^n_X(\delta)$. Proposition \ref{prop:convnonparam} states a subsequent result for the nonparametric estimates $\hat{p}^n_l$. 

For Proposition \ref{prop:convparam} we will need inequalities (\ref{eq:wh}) and (\ref{eq:logstable}). These express stability results for the eigenvalues of Hermitian matrices and for the Hermitian matrix function $\Log$. Let $A$ and $B$ be Hermitian $d \times d$ matrices, for some $d \geq 1$. For $1 \leq i \leq d$ let $\alpha_i$ and $\beta_i$ be the eigenvalues of $A$ and $B$ respectively. Suppose they are arranged in nondecreasing order. We have
\begin{equation} \label{eq:wh}
 \sum^d_{i = 1} (\beta_i - \alpha_i)^2 \leq |B-A|^2
\end{equation}
where $|.|$ is the Euclidean matrix norm. This inequality is known as the Wielandt-Hoffman theorem. In~\cite{vlg:89}, it is stated for $A$ and $B$ real symmetric. The general case of Hermitian $A$ and $B$ can be obtained from this statement using a canonical realification isomorphism.  

Suppose $A$ and $B$ are positive definite. For our purpose it is suitable to assume both $\lambda(A)$ and $\lambda(B)$ are contained in an interval $[k,1]$ for some $k > 0$. Under this assumption we have the following Lipschitz property 
\begin{equation} \label{eq:logstable}
 |\Log(B) - \Log(A)| \leq \sqrt{d}k^{-2} |B-A|
\end{equation}
In order to obtain (\ref{eq:logstable}) it is possible to start by expressing $\Log(A)$ as follows 
$$
\Log(A) = \int^1_0 (A-I_d)[t(A-I_d) + I_d]^{-1} dt
$$
This expression results from a similar one for the real logarithm applied to each eigenvalue of $A$. Subtracting the same expression for $\Log(B)$, (\ref{eq:logstable}) follows by simple calculations. 
\begin{proposition} \label{prop:convparam}
 For all $\delta \in \Irr(G)$ we have the limit in probability $ \lim_n \hat{\phi}^n_X(\delta) = \phi_X(\delta)$.
\end{proposition}
\begin{proof} We only need to consider $\delta \neq \delta_0$. Indeed, $\hat{\phi}^n_X(\delta_0) = \phi_X(\delta_0) = 1$ for all $n \geq 1$. Let $\delta \neq \delta_0$, for all $n \geq 1$ we have
$$
|\hat{\phi}^n_Z(\delta)|_{op} \leq \frac{1}{2n} \sum^n_{m=1} |U^\delta(Z_m)|_{op} + |U^\delta(Z_m)^\dagger|_{op} =  1
$$
where $|.|_{op}$ is the operator matrix norm. Passing to the limit, we have the same inequality for $\phi_Z(\delta)$. It follows that all eigenvalues of $\hat{\phi}^n_Z(\delta)$ or $\phi_Z(\delta)$ are $\leq 1$. Since $\phi_Z(\delta)$ is positive definite, there exists $k_\delta > 0$ such that $\lambda(\phi_Z(\delta)) \subset [k_\delta,1]$. For $n \geq 1$, note $\tilde{R}^n_\delta$ the event 
$$
\tilde{R}^n_\delta = \lbrace \lambda(\hat{\phi}^n_Z(\delta)) \subset [k_\delta/2,1] \rbrace 
$$
From inequality (\ref{eq:wh}) we have
$$
\mathbb{P}(\Omega - \tilde{R}^n_\delta) \leq \mathbb{P}( |\hat{\phi}^n_Z(\delta) - \phi_Z(\delta)| > k_\delta/2)
$$
Since $\tilde{R}^n_\delta \subset R^n_\delta$, it follows from inequality (\ref{eq:logstable}) that
$$
\mathbb{P}(|\hat{\phi}^n_X(\delta) - \phi_X(\delta)|> \varepsilon \cap \tilde{R}^n_\delta) \leq 
\mathbb{P}(|\hat{\phi}^n_Z(\delta) - \phi_Z(\delta)|> k^2_\delta \varepsilon/L)
$$
for all $\varepsilon > 0$, where $L = 4\sqrt{d_\delta}/T\lambda$. 

The proof can be completed by a usual application of Chebychev's inequality, 
\begin{equation} \label{eq:pconvparam}
\mathbb{P}(|\hat{\phi}^n_X(\delta) - \phi_X(\delta)|> \varepsilon) \leq \left(\frac{8 + 2L^2/\varepsilon^2}{n} \right) \left(\frac{\sqrt{d_\delta}}{k^2_\delta}\right)^2 
 \end{equation}
for all $\varepsilon > 0$.
\end{proof}
Proposition \ref{prop:convnonparam} relies on Proposition \ref{prop:convparam} and the Peter-Weyl theorem. It implies the existence of sequences $(\hat{p}_k)_{k \geq 1}$, of nonparametric estimates given by (\ref{eq:nonparam}), converging to $p$ in probability in $L^2(G,\C)$ with any prescribed rate of convergence. Convergence in probability in $L^2(G,\C)$ means that the following limit in probability holds
$$
\lim_k \Vert \hat{p}_k - p \Vert = 0
$$
where $\Vert.\Vert$ is the $L^2(G,\C)$ norm. It is clear from (\ref{eq:nonparam}) that for all $k \geq 1$ we have $\hat{p}_k \in L^2(G,\C)$. In order to obtain nonparametric estimators in $L^2(G,\R)$ and converging to $p$ in the same sense, it is enough to consider the real parts of the $\hat{p}_k$. The following proof of Proposition \ref{prop:convnonparam} implicitly uses Plancherel's formula as in~\cite{kk:08}. 
\begin{proposition} \label{prop:convnonparam}
Putting $K = 0$ in (\ref{eq:nonparam}), we have the limit in probability
$$
\lim_l \lim_n \Vert \hat{p}^n_l - p \Vert = 0
$$
\end{proposition}
\begin{proof}
 For $l \geq 1$ let $p^{}_l \in L^2(G,\C)$ be given by
$$
p^{}_l(g) = 1 + \sum_{\delta \in \Gamma_l} \tr\left( \phi_X(\delta) U^\delta(g)^\dagger \right)
$$
for $g \in G$. By the Peter-Weyl theorem, $\lim_l \Vert p^{}_l - p \Vert = 0$. By (\ref{eq:nonparam}) and Proposition \ref{prop:convparam} we have $\lim_n \Vert \hat{p}^n_l - p^{}_l \Vert = 0$ in probability for all $l \geq 1$. The proposition follows by observing that 
\begin{equation} \label{eq:plancherel}
\Vert \hat{p}^n_l - p \Vert^2 = \Vert \hat{p}^n_l - p^{}_l \Vert^2 + \Vert p^{}_l - p \Vert^2 
\end{equation}
for all $n,l \geq 1$. \end{proof}
Proposition \ref{prop:convparam} obtained convergence in probability of the parametric estimates $\hat{\phi}^n_X(\delta)$ for all $\delta \in \Irr(G)$. These parametric estimates depend only on the observations. In particular, they can be evaluated without any \textit{a priori} knowledge of $p$. By introducing such knowledge, it is possible to define parametric estimates $\tilde{\phi}^n_X(\delta)$ converging in the square mean to the same limits $\phi_X(\delta)$. For $\delta \in \Irr(G)$ and $n \geq 1$ the $\tilde{\phi}^n_X(\delta)$ are given by
$$
\begin{array}{rclrl}
\tilde{\phi}^n_X(\delta) & = & 0 &  \text{on} &  \Omega - \tilde{R}^n_\delta \\
\tilde{\phi}^n_X(\delta) & = & \frac{1}{T\lambda} \Log\left[\hat{\phi}^n_Z(\delta)\right] + 
                             \left(\bar{\lambda}/\lambda\right) I_{d_\delta} & \text{on} &    \tilde{R}^n_\delta 
\end{array}
$$ 
where the events $\tilde{R}^n_\delta$ are as in the proof of Proposition \ref{prop:convparam} and we assume known \textit{a priori} constants $k_\delta$ necessary for their definition. As in (\ref{eq:nonparam}), we can define nonparametric estimates $\tilde{p}^n_l$ where for $n,l \geq 1$
$$
\tilde{p}^n_l(g) = 1 + \sum_{\delta \in \Gamma_l} f_\delta \tr\left( \tilde{\phi}^n_X(\delta) U^\delta(g)^\dagger \right) \;\;\;g \in G
$$
For all $\delta \in \Irr(G)$ and $n \geq 1$ we have 
\begin{equation} \label{eq:l2convparam}
\mathbb{E}| \tilde{\phi}^n_X(\delta) - \phi_X(\delta)|^2 \leq \frac{L^\prime}{n} \left(\frac{d_\delta}{k^2_\delta}\right)^2
\end{equation}
where $L^\prime$ is a constant depending on the product $T\lambda$. This follows by a reasoning similar to the proof of Proposition \ref{prop:convparam}. Moreover, for all $n,l \geq 1$ we have after putting $K = 0$ 
\begin{equation} \label{eq:l2plancherel}
\mathbb{E}\Vert \tilde{p}^n_l - p \Vert^2 \leq  
\frac{L^\prime}{n}\sum_{\delta \in \Gamma_l} (d^3_\delta/k^4_\delta) + \Vert p^{}_l - p \Vert^2
\end{equation}
for the functions $p^{}_l$ defined in the proof of Proposition \ref{prop:convnonparam}. This follows from Plancherel's formula in (\ref{eq:plancherel}).

We have characterized the convergence of parametric estimates using (\ref{eq:pconvparam}) and (\ref{eq:l2convparam}) and the convergence of nonparametric estimates using (\ref{eq:plancherel}) and (\ref{eq:l2plancherel}). We make the following remarks on these formulae. Inequalities (\ref{eq:pconvparam}) and (\ref{eq:l2convparam}) only give gross bounds for the rate of convergence of parametric estimates. The quality of these bounds improves when the constants $k_\delta$ are greater, \textit{i.e.} closer to the value $1$. This is equivalent to the $L^2(G,\R)$ distance between $p$ and the uniform density being greater. This last point can be appreciated in relation to the example of figure \ref{fig:uniformization} in \ref{subsec:simu}.

(\ref{eq:plancherel}) and (\ref{eq:l2plancherel}) describe the convergence of nonparametric estimates in a way similar to the one used in~\cite{kk:08}. Indeed, the nonparametric estimation error is decomposed into two terms. One is given by the parametric estimation error and the other depends only on $p$. This second term is given by the convergence of the Fourier series of $p$. This is determined by the smoothness properties of $p$. We note the two following differences with~\cite{kk:08}, both related to the indirect nature of our observations. First, the first and second terms in (\ref{eq:l2plancherel}) can not be identified as the "variance" and "bias" of $\tilde{p}^n_l$. Second, (\ref{eq:l2plancherel}) characterizes the nonparametric estimation error as depending on the whole spectrum of $p$ --through the constants $k_\delta$-- rather than just its smoothness properties. 

We finally return to the role of the parameter $K$ introduced in (\ref{eq:nonparam}). For simplicity, we have put $K = 0$ for Proposition \ref{prop:convnonparam} and inequality (\ref{eq:l2plancherel}). Let $K > 0$. The following function $p_K \in L^2(G,\R)$ is an infinitely differentiable probability density~\cite{liao,kk:08}
\begin{equation} \label{eq:smoothedp}
p_K(g) = 1 + \sum_{\delta \neq \delta_0} f_\delta \tr(A_\delta U^\delta(g)^\dagger)
\end{equation}
Using the same $K$ in (\ref{eq:nonparam}) and proceeding as for proposition \ref{prop:convnonparam} it is possible to obtain the limit in probability
$$
\lim_l \lim_n \Vert \hat{p}^n_l - p_K \Vert = 0
$$  
A similar limit also holds for the $\tilde{p}^n_l$. Note that in addition to being smooth, $p_K$ can be chosen arbitrarily close to $p$ in $L^2(G,\R)$ for $K > 0$ small enough. 

\section{Decompounding on $SO(3)$ and multiple scattering} \label{sec:so(3)}
This section fulfills two goals. First, it summarizes recent use of compound Poisson processes on the rotation group $SO(3)$ in the modelling of multiple scattering and introduces decompounding on $SO(3)$ as a physical inverse problem. Second, it illustrates the characteristic function method presented in \ref{subsec:estimators} by applying it to a numerical example of decompounding on $SO(3)$. nonparametric estimation on the rotation group $SO(3)$ has received special attention~\cite{kim:98,kk:02}. It is important to many concrete applications and constitutes a privileged starting point for generalization to compact groups. 

\subsection{The compound Poisson model for multiple scattering} \label{subsec:scattering}
Many experimental and applied settings aim to infer the properties of complex, \textit{e.g.} geophysical or biological, media by considering multiple scattering of mechanical or electromagnetic waves by these media. Inference problems arising in this way are formulated as physical inverse problems within the framework of various approximations of the exact equations of radiative transfer. See~\cite{sato:98,xu:02,sheng:95}. 

A compound Poisson model for the direct problem of multiple scattering was considered by Ning \textit{et al.}~\cite{n:95}. It is based on a $\R$-valued compound Poisson process. Consideration of compound Poisson processes on $SO(3)$ leads to a model of multiple scattering which is sufficiently precise as well as amenable to statistical treatment. This model extends the validity of the small angles approximation of radiative transfer. It also allows the formulation of the physical inverse problem of multiple scattering as a statistical nonparametric estimation problem. 

We give an example expanding the above discussion. The development of Section \ref{sec:cpp} is converted into the terminology of radiative transfer, see~\cite{ish:78}. Certain usual results in harmonic analysis on $SO(3)$ are here referred to freely. They are set down in a precise form in \ref{subsec:harso(3)}. 

A scalar plane wave is perpendicularly incident upon a plane parallel multiple scattering layer of thickness $H$. The velocity of the wave in the layer is normalized so that we have $\tau = \ell$ for the mean free time $\tau$ and mean free path $\ell$. Coordinates and time origin are chosen so that the wave enters the layer at time $0$ with direction of propagation $s(0) = (0,0,1)$. After time $t$ in the layer this direction of propagation becomes $s(t) = (s^1(t),s^2(t),s^3(t))$. This is considered to be a random variable with values on the unit sphere $S^2 \subset \R^3$. The distribution of the random variable $s(H)$ is noted $I_H$. It is identified with the normalized angular pattern of intensity transmitted by the layer. We return below to the validity of this identification. 

The interaction of the wave with the layer takes place in the form of a succession of scattering events. These are understood as interaction of the wave with individual scatterers present at random emplacements throughout the layer. The random number of scattering events up to time $0 \leq t \leq H$ will be noted $N(t)$. Suppose the $n^{th}$ scattering event takes place at the time
$0 \leq T_n \leq H$. This affects the direction of propagation as follows
\begin{equation} \label{eq:scaterevent}
s(T_n) = s(T_n-)X_n
\end{equation}
Here $X_n$ is a random variable with values in $SO(3)$. It is identified with a random orthogonal matrix. Formula
(\ref{eq:scaterevent}) is understood as a matrix equality where $s(T_n)$ and $s(T_n-)$ are line vectors. From (\ref{eq:scaterevent}) and the definition of $N(t)$ we can write for $0 \leq t \leq H$
\begin{equation} \label{eq:multiplescater}
s(t) = s(0) \left( \prod^{N(t)}_{n=0} X_n \right)
\end{equation}
A certain number of standard physical hypotheses can be replaced in (\ref{eq:multiplescater}). This will allow for the random product therein to be exhibited as a conjugate invariant compound Poisson process on $SO(3)$. 

Under the condition $\ell \ll H$ it is possible to make the hypothesis that the time between successive scattering events has an exponential distribution~\cite{sheng:95}. This allows us to model $N(t)$ as a Poisson process with parameter $1/\ell$. Moreover, we suppose the scatterers identical and scattering events independent. This amounts to taking the $SO(3)$-valued random variables $X_n$ to be \textit{i.i.d.}. If the additional assumption is accepted that the number of scattering events is independent of the whole outcome of these events then formula (\ref{eq:multiplescater}) can be rewritten $0 \leq t \leq H$
\begin{equation} \label{eq:multiplescaterCPP}
s(t) = s(0) Y(t)
\end{equation} 
Where $Y$ is a (left) compound Poisson process on $SO(3)$ with parameter $1/\ell$. It is usual to assume that the random variables $X_n$ have a common probability density $p$. For homogeneity with \ref{sec:decomp} we mention that $p$ is a square integrable probability density with respect to the Haar measure of $SO(3)$. In the theory of radiative transfer, $p$ is known as the phase function of the layer~\cite{ish:78}. 

In order to simplify the Fourier series of $p$ to a Legendre series (\ref{eq:lserso3}) we profit from the physical hypothesis of statistical isotropy. This implies that scattering events in the layer as given by (\ref{eq:scaterevent}) are symmetric around the direction of propagation $s(T_n-)$. Statistical isotropy is a valid assumption in a plurality of concrete situations. It is verified by analytical models such as Gaussian and Henyey-Greenstein phase functions, commonly used to describe scattering in geophysical and biological media~\cite{klimes:02}. 

Under the hypothesis of statistical isotropy the phase function $p$ is a zonal function in the sense precised in \ref{subsec:harso(3)}. It admits a Legendre series (\ref{eq:lserso3}) wherein the coefficients $a_\delta$ for $\delta \in \N$ are said to form the associated power spectrum of heterogenities~\cite{ish:78}. If $p$ is the Henyey-Greenstein phase function then the power spectrum of heterogenities is given by $a_\delta = g^\delta$ for $\delta \in \N$ and $p$ can be expressed in the closed form~\cite{klimes:02,kokh:97}
\begin{equation} \label{eq:HG}
 p(\cos \theta) = \frac{1-g^2}{(1+g^2-2g \cos \theta)^\frac{3}{2}}
\end{equation}
In this formula the variable $\theta \in [0,\pi]$ refers to the scattering angle from an individual scatterer. It is given a mathematical definition in formula (\ref{eq:lserso3}) of \ref{subsec:harso(3)}. The parameter $g \in [0,1[$ is called the anisotropy or asymmetry parameter. It can be shown to give the average cosine of the scattering angle $\theta$. For the scattering of light waves by water clouds and blood we have respectively $g = 0.85$ and $g = 0.95$, see~\cite{kokh:97}.  

Proposition \ref{prop:char} of Section \ref{sec:cpp} can be used to give the angular pattern of transmitted intensity $I_H$ in terms of the power spectrum of heterogenities. This is expressed in the following equation (\ref{eq:transferl}). This relates the directly observable outcome of multiple scattering in the layer to the constitutive microscopic properties of the layer, typically quite difficult to ascertain directly. Replacing in Proposition \ref{prop:char} the definition of the process $Y$ of (\ref{eq:multiplescaterCPP}) and using the Legendre series (\ref{eq:lserso3}) of $p$ we have
\begin{equation} \label{eq:transferl}
 \frac{I_H(\theta)}{2\pi} = \sum_{\delta \geq 0} (2 \delta + 1) e^{\frac{H}{\ell}\left( a_\delta - 1 \right)} 
                       \int^\theta_0 P_\delta(\cos \xi)\sin \xi d\xi
\end{equation}
For the ratio $I_H(\theta)$ of intensity transmitted within a pencil of angle $2\theta$ around $s(0)$.  

Equation (\ref{eq:transferl}) is well known in the small angles approximation of radiative transfer where it is derived under the assumption of strong forward scattering~\cite{ish:78}. Mathematically, this translates into a phase function $p$ with a sharp peak around $\theta = 0$. Our probabilistic development of equation (\ref{eq:transferl}) does not explicitly make this assumption. However, the identification of $I_H$ with the angular pattern of transmitted intensity implicitly requires for all the intensity of the wave entering the layer to be transmitted. This precludes an important deviation between $s(0)$ and $s(H)$.

Equation (\ref{eq:transferl}) is an interesting starting point for the formulation of the physical inverse problem of multiple scattering. Supposing a situation where this equation holds, being able to invert it implies access to the power spectrum of heterogenities or alternatively the phase function from direct intensity measurements. This implies inference of physical parameters such as the parameter $g$ of the Henyey-Greenstein phase function or determination of microscopic properties such as the shape of individual scatterers~\cite{kokh:97}. 

Our use of compound Poisson processes on $SO(3)$ to model multiple scattering lead to the probabilistic counterpart (\ref{eq:multiplescaterCPP}) of equation (\ref{eq:transferl}). In relation to (\ref{eq:multiplescaterCPP}), the physical inverse problem inherent to equation (\ref{eq:transferl}) is reformulated as a statistical estimation problem. This appears as the problem of
decompounding on $SO(3)$ or some related parametric estimation problem. A crucial difference between the two approaches is that they proceed from different types of data. 

Suppose the distribution of $s(0)$ is known and symmetric around $(0,0,1)$ --this is the case in many experimental settings. Instead of carrying out measurements of transmitted intensity, it is possible to make observations of $s(H)$. Under the hypothesis of statistical isotropy these observations of $s(H)$ are equivalent to observations of $Y(H)$. If our objective is to estimate the phase function $p$ then we have to deal with decompounding on $SO(3)$ from low frequency observations of $Y$. In many cases, we could be interested in the power spectrum of heterogenities or some related physical parameters. We then have to deal with a parametric estimation problem. 

\subsection{Harmonic analysis on $SO(3)$} \label{subsec:harso(3)} 
We here make a short digression on harmonic analysis on $SO(3)$ in order to clarify the references made to this subject in \ref{subsec:scattering} and to prepare for \ref{subsec:simu}. $SO(3)$ is often used as the archetype compact connected Lie group. Essentially, we will specify the Peter-Weyl theorem as stated in Section \ref{sec:hap} to the case $G = SO(3)$. For the following see~\cite{kk:02} or the more detailed account in~\cite{ck}. 

We use the notation of Section \ref{sec:hap}. In particular, $\mu$ denotes the Haar measure of $SO(3)$. It is possible to identify $\Irr(SO(3)) = \N$ so that $d_\delta = 2\delta + 1$ for each $\delta \in \Irr(SO(3))$. With this identification, the most current choice of functions $U^\delta :SO(3) \rightarrow SU(d_\delta)$ can be given in analytical form using the parameterization of $SO(3)$ by Euler angles.

The $ZYZ$ Euler angles $\varphi,\psi \in [0,2\pi]$ and $\theta \in [0,\pi]$ are well defined coordinates only on a subset of $SO(3)$. This is however a dense subset in the Euclidean topology of $SO(3)$ and has Haar measure equal to $1$. Let $p :SO(3) \rightarrow \C$. If $p$ is continuous or $p \in L^2(SO(3),\C)$ it follows that $p$ can be identified with a function of the Euler angles $p \equiv p(\varphi,\theta,\psi)$. The chosen functions $U^\delta$ are extended by continuity from the following expression for their matrix elements
\begin{equation} \label{eq:wigner}
U^\delta_{ab}(\varphi,\theta,\psi)= e^{{-\tt i}a \varphi}d^\delta_{ab}(\cos \theta)e^{{-\tt i} b\psi}
\end{equation}
for $\delta \in \Irr(SO(3))$ and $-\delta \leq a,b \leq \delta$. The notation $d^\delta_{ab}$ is used for the real-valued Wigner d-functions, which can be given in terms of the Jacobi polynomials. For $\delta \in \Irr(SO(3))$ we have $d^\delta_{00} = P_\delta$ the Legendre polynomial of order $\delta$. 

The Haar measure $\mu$ is expressed in the coordinates $(\varphi,\theta,\psi)$ as follows 
$$
d\mu(\varphi,\theta,\psi) = \frac{1}{8\pi^2} \sin\theta d\varphi d\theta d\psi
$$
Suppose a function $p \in L^2(SO(3),\C)$ is expressed in the form $p(\varphi,\theta,\psi)$. In order to obtain its Fourier coefficients, it is enough to replace the above expressions for the functions $U^\delta$ and $\mu$ in formula (\ref{eq:fcoef}). This formula then reduces to a triple integral. By the Peter-Weyl theorem, the Fourier coefficients of $p$ give rise to a Fourier series approximating $p$ in $L^2(SO(3),\C)$.

The class of zonal functions on $SO(3)$ arises in relation to the hypothesis of statistical isotropy mentioned in \ref{subsec:scattering}. We will say that a function $p \in L^2(SO(3),\C)$ is zonal if $p \equiv p(\theta)$. That is, if the expression of $p$ in the coordinates $(\varphi,\theta,\psi)$ depends only on $\theta$. Zonal functions form a closed subspace of $p \in L^2(SO(3),\C)$. If $p$ is a zonal function then its Fourier series reduces to a Legendre series
\begin{equation} \label{eq:lserso3}
 p(\theta) = \sum_{\delta \geq 0} (2\delta+1)a_\delta P_\delta(\cos \theta) 
\end{equation}
where for $\delta \geq 0$ the Legendre coefficient $a_\delta$ is given by
\begin{equation} \label{eq:lcoef}
 a_\delta = \frac{1}{2} \int^\pi_0 p(\theta) P_\delta(\cos \theta) \sin \theta d\theta
\end{equation}
Identities (\ref{eq:lserso3}) and (\ref{eq:lcoef}) can be found as follows. Let $p$ be a zonal function. For $\delta \in \Irr(SO(3))$ let $A_\delta$ be the Fourier coefficients of $p$ obtained by replacement in (\ref{eq:fcoef}). The matrix elements of each $A_\delta$ are noted $A^{ab}_\delta$ for $-\delta \leq a,b \leq \delta$. For all $\delta,a,b$ as above we have that $A^{ab}_\delta$ is given by --this follows using (\ref{eq:fcoef})
$$
\frac{1}{8\pi^2} \int^{2\pi}_0 \int^\pi_0 \int^{2\pi}_0 e^{{\tt i}b \varphi} p(\theta) d^\delta_{ba}(\cos \theta)e^{{\tt i} a\psi}
                                       \sin\theta d\varphi d\theta d\psi
$$
Thus for all $\delta \in \Irr(SO(3))$ we have that $A^{ab}_\delta \neq 0$ only if $a=b=0$. In other words the matrix $A_\delta$ contains at most one nonzero element. This is the diagonal element $A^{00}_\delta = a_\delta$ given by identity (\ref{eq:lcoef}).
Identity (\ref{eq:lserso3}) follows by constructing the Fourier series of $p$ as in (\ref{eq:fseries}). 

\subsection{Numerical simulations}\label{subsec:simu}
Here we will illustrate the characteristic function method of \ref{subsec:estimators} by applying it to a numerical example of decompounding on $SO(3)$. Within this example we will consider a parametric estimation problem related to a physical inverse problem as described in \ref{subsec:scattering}. Our example is of a compound Poisson process $Y$ on $SO(3)$. As in \ref{subsec:scattering}, $SO(3)$-valued random variables are identified with random orthogonal matrices. For $t \geq 0$, 
$$
Y(t) = \prod^{N(t)}_{n=0} X_n
$$
where the Poisson process $N$ has parameter $\lambda=0.3$ and the random variables $X_n$ have a common probability density $p$ given by expression (\ref{eq:HG}). Four values will be considered for the parameter $g$ in this expression: $0.85, 0.9, 0.95$ and $0.99$. We will put $T = 10$. We simulate a number $n$ of \textit{i.i.d.} observations of $Y(T)$. The following values of $n$ are used: $500, 5000$ and $50000$. Note that on average the number $N(T)$ of factors involved in the random product $Y(T)$ is equal to $3$. 

\begin{figure}[b!]
\begin{center}
\subfigure[Histogram of $\cos \theta$ under density $p$]
{\centering
{\includegraphics[height=5cm,width=8cm]{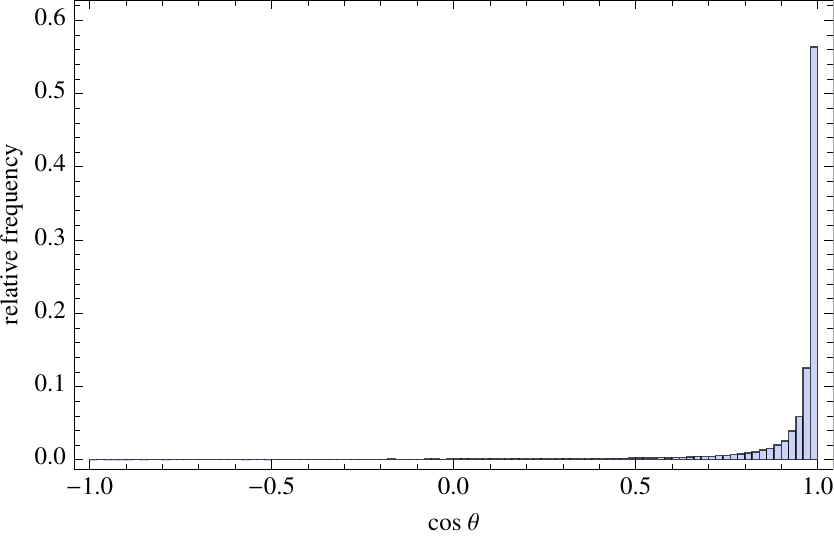}
\label{fig:a}
}}
\subfigure[Histogram of $\cos \theta$ under distribution of $Y(T)$]
{\centering
{\includegraphics[height=5cm,width=8cm]{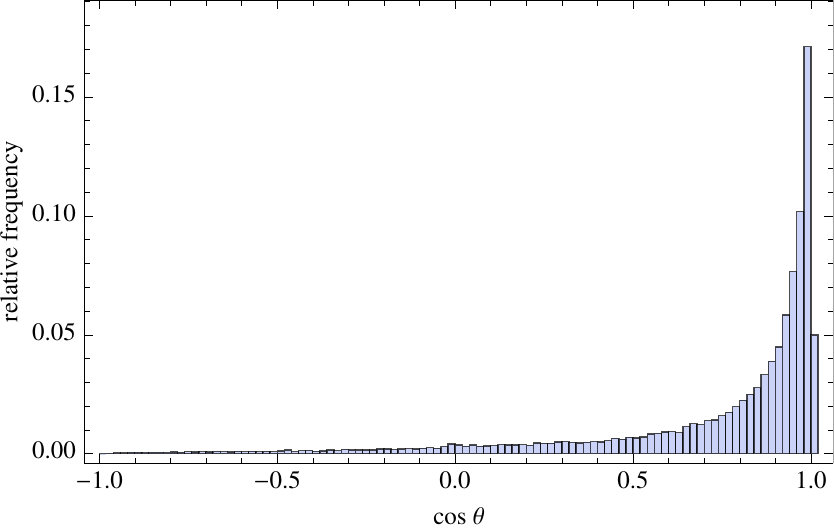}
\label{fig:b}}}
\caption{Compounding transformation of $p$ (histograms)}
\end{center}
\label{fig:compounding}
\end{figure}

Before going on, we confirm that the method of \ref{subsec:estimators} can be applied for this example. In other words, that the $X_n$ with the proposed density $p$ are inverse invariant. This follows from the development after identities (\ref{eq:lserso3}) and (\ref{eq:lcoef}). Indeed, the matrices $A_\delta$ obtained for $p$ are diagonal with exactly one nonzero diagonal element $a_\delta =g^\delta$. Since $g$ is real we have that $A_\delta$ is Hermitian for all $\delta \in \Irr(SO(3))$. Inverse invariance follows by \ref{sym:a} of Proposition \ref{prop:sym}. 

\begin{figure}[t!]
\begin{center}
\subfigure[Estimated Legendre coefficients $\hat{a}^n_\delta$ from decompounding]
{\centering
{\includegraphics[height=5cm,width=8cm]{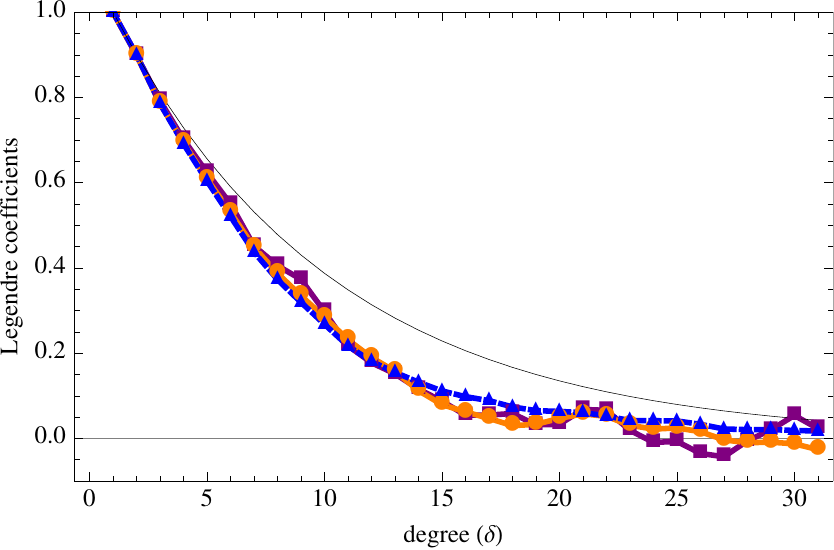}
 \label{fig:c}
}}
\subfigure[Corresponding estimates $\hat{g}^n_\delta$ of $g$ (anisotropy parameter)]
{\centering
{\includegraphics[height=5cm,width=8cm]{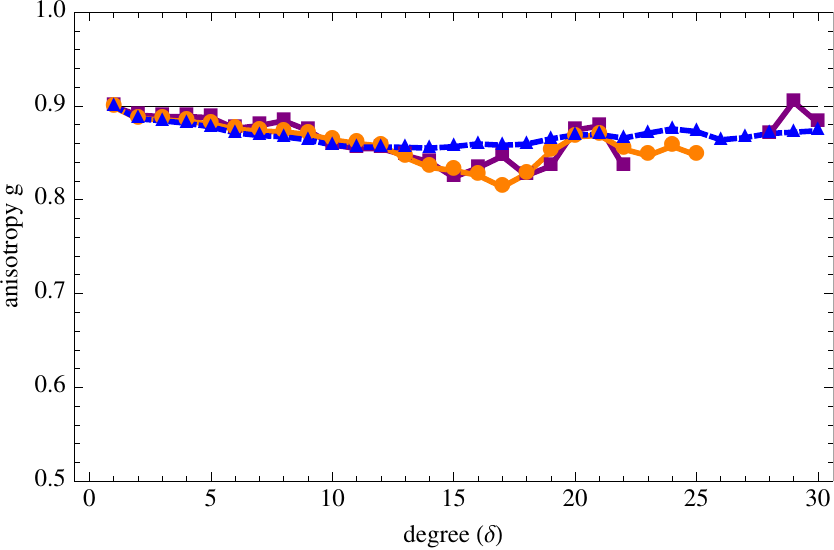}
\label{fig:d}}}
\caption{Influence of $n$ ($\square=5*10^2$; $\circ=5*10^3$; $\triangle=5*10^4$)}
\end{center}
\label{fig:decompounding}
\end{figure}

\begin{figure}[t!]
\begin{center}
{\centering
{\includegraphics[height=5cm,width=8cm]{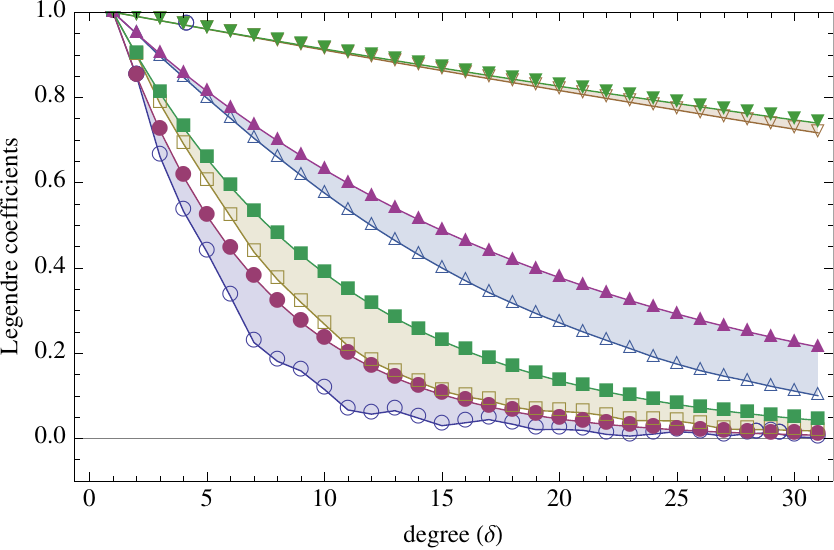}
}}
\caption{Influence of $g$ ($\circ=0.85$; $\square=0.9$; $\triangle=0.95$; $\nabla=0.99$)}
\end{center}
\label{fig:uniformization}
\end{figure}

We will present three sets of figures. Figure \ref{fig:compounding} is concerned with the compounding transformation of $p$. Figure \ref{fig:decompounding} illustrates the influence of $n$ on parametric and nonparametric estimation errors. Figure \ref{fig:uniformization} studies the influence of $g$ on the nonparametric estimation error for fixed $n$. For figures \ref{fig:compounding} and \ref{fig:decompounding} we have $g=0.9$. For figures \ref{fig:compounding} and \ref{fig:uniformization} we have $n=50000$. We now comment on each of these figures.

Figure \ref{fig:compounding} illustrates the relation between the distribution of the $X_n$ as given by the density $p$ and the distribution of $Y(T)$. Both these distributions are studied using histograms. The histogram in figure \ref{fig:a} is for the cosine of the Euler angle $\theta \in [0,\pi]$ associated with the random variable $X_1$. The histogram in figure \ref{fig:b} is for the cosine of $\theta$ associated with $Y(T)$. 

Figure \ref{fig:compounding} is concerned with the direct compounding transformation rather than the inverse decompounding transformation. It is meant to show the histogram in figure \ref{fig:b} as function of the one in \ref{fig:a}. As expected, the latter histogram appears as a wider version of the former. This corresponds to the content of Proposition \ref{prop:uniformize} of Section \ref{sec:cpp}. Note also that the dominant value in figure \ref{fig:b} has moved away from $\theta = 0$.

For figure \ref{fig:decompounding}, the observations made of $Y(T)$ are used to carry out the decompounding approach of \ref{subsec:estimators}. Parametric and nonparametric estimation errors are given graphically for different values of $n$. Figure \ref{fig:c} compares the estimated Legendre coefficients of $p$ to their theoretical values $a_\delta=g^\delta$ for $\delta \geq 0$. In figure \ref{fig:d}, \textit{a priori} knowledge of the analytical form of the $a_\delta$ is supposed. This is used to estimate $g$. A different parametric estimate is obtained from each estimated Legendre coefficient. In figures \ref{fig:c} and \ref{fig:d} theoretical values are represented by a solid line.

In figure \ref{fig:c} we have the estimated first $l=31$ Legendre coefficients for each value of $n$. Let us note these coefficients $\hat{a}^n_\delta$ for $0 \leq \delta \leq l$ and the corresponding value of $n$. They can be used to evaluate a nonparametric estimate of $p$ as in formula (\ref{eq:nonparam}). This is done by replacing them in a truncated Legendre series (\ref{eq:lserso3}). We have the nonparametric estimate of $p$ which we note $\hat{p}^n_l$
$$
\hat{p}^n_l(\theta) = 1 + \sum^{l-1}_{\delta = 1} (2\delta+1) \hat{a}^n_\delta P_\delta(\cos \theta)
$$
where for all values of $n$ we have that $\hat{a}^n_0=a_0=1$. Depending on $n$, the random nonparametric estimation error from $\hat{p}^n_l$ is given by 
$$
\sum_{\delta < l} (2\delta+1) (\hat{a}^n_\delta - a_\delta)^2 + \sum_{\delta \geq l} (2\delta+1) a^2_\delta
$$
this is the squared $L^2(SO(3),\R)$ distance between $\hat{p}^n_l$ and $p$. In figure \ref{fig:c} the sum over $\delta < l$ appears as a weighted quadratic deviation between estimated and theoretical values.  

In figure \ref{fig:d} the estimates $\hat{a}^n_\delta$ are used to give naive estimates $\hat{g}^n_\delta$ of $g$ based on the analytical form of the $a_\delta$. The error in each of these estimates $\hat{g}^n_\delta$ is directly related to the error in the estimate $\hat{a}^n_\delta$. This latter error is shown for each $\delta$ and $n$ in figure \ref{fig:c}. The influence of $n$ is not important for small values of $\delta$. Visually, the $\hat{a}^n_\delta$ in figure \ref{fig:c} agree independently of $n$ for $0 \leq \delta \leq 5$. For $n=50000$ the $\hat{a}^n_\delta$ appear to have a regular dependence on $\delta$. For $n=5000$ and $n=500$ we have an irregular dependence of the $\hat{a}^n_\delta$ on $\delta$, especially for $\delta \geq 20$. Moreover, for $\delta \geq 25$ we have negative values of $\hat{a}^n_\delta$, clearly inconsistent with the form $a_\delta=g^\delta$. These values do not allow the evaluation of corresponding parametric estimates $\hat{g}^n_\delta$.

Let us remind that $g$ is an important parameter in multiple scattering applications. For multiple scattering media with Henyey-Greenstein phase function (\ref{eq:HG}), $g$ is the main parameter characterizing the scattering process. Its estimation from observations as the ones described in \ref{subsec:scattering} is equivalent to a physical inverse problem. This leads to the physical interpretation of the parametric estimation problem represented in figure \ref{fig:d}. 

For figure \ref{fig:uniformization} we have $n=50000$. For each value of $g$ we simulated $n$ observations of $Y(T)$ and calculated estimates of the Legendre coefficients of $p$ as for figure \ref{fig:c}. Estimated and theoretical Legendre coefficients are respectively represented by empty and filled in symbols. It is clear from this figure that the nonparametric estimation error is smaller for larger values of $g$. Estimation of the Legendre coefficients is virtually exact for $g= 0.99$.

In order to understand this behavior we note that $g$ in (\ref{eq:HG}) gives the concentration of $p$ near the value $\theta=0$. Indeed, when $g=0$ the function $p$ is constant and the random variables $X_n$ are uniformly distributed on $SO(3)$. In the limit $g \uparrow 1$ we have that each random variable $X_n$ is almost surely equal to the identity matrix. Conditionally on the event $\lbrace N(T) > 0\rbrace$, the distribution of $Y(T)$ is a mixture of distributions with Henyey-Greenstein density. More precisely, for all $ n > 0$ we have the conditional probability density for the Euler angle $\theta$ associated with $Y(T)$
$$
p(\theta|N(T)=n) = \frac{1-g^{2n}}{(1+g^{2n}-2g^n \cos \theta)^\frac{3}{2}} 
$$
In particular, in the limit $g \uparrow 1$  we have that $Y(T)$ is almost surely equal to the identity matrix. Conditionally on $\lbrace N(T) > 0\rbrace$, we have in the limit $g \downarrow 0$ that $Y(T)$ is uniformly distributed on $SO(3)$. 

Let us note that in our example $\mathbb{P}(N(T) > 0) \simeq 0.96$. Figure \ref{fig:uniformization} can be understood in light of the above discussion. For greater values of $g$, observations of $Y(T)$ are concentrated near the identity matrix. This leads to fast convergence of our estimates for the Legendre coefficients of $p$. For smaller values of $g$, observations of $Y(T)$ are more dispersed and the convergence of estimates is slower. In the limit $g \downarrow 0$ the observations are close to uniformly distributed on $SO(3)$ and our approach breaks down due to numerical problems.  

\section{Conclusion} \label{sec:conclusions}
Nonparametric estimation on compact Lie groups, especially using characteristic function methods, is by now a relatively familiar topic in relation to several engineering applications. It has received comprehensive treatment in the case where estimation is carried out directly from some stationary process. That is, from \textit{i.i.d.} observations of a group-valued random variable. This paper has applied a characteristic function method to the problem of decompounding on compact Lie groups. For this problem, nonparametric estimation is required from indirect observations defined in terms of a nonstationary process.  

A first approach of decompounding on compact Lie groups was given. It was guided by existing characteristic function methods for the classical problem of decompounding. These methods were transposed directly to the setting of harmonic analysis on compact Lie groups. Under a suitable symmetry hypothesis, treatment of the indirect nature of observations was simplified. The ensuing nonparametric estimation error was characterized as depending on the whole spectrum of the target density rather than just its smoothness class. In some aspects, our approach of decompounding on compact Lie groups might appear summary. We hope however that is will attract attention to various problems of the statistics of nonstationary stochastic processes on groups.  

This paper also discussed the importance of decompounding on $SO(3)$ to the physical inverse problem of multiple scattering. Under a probabilistic interpretation of the theory of radiative transfer, models based on compound Poisson processes on $SO(3)$ were found consistent with the results of the small angles approximation of radiative transfer. The possibility of reformulating physical inverse problems of multiple scattering as parametric or nonparametric statistical estimation problems was discussed. The statistical nature of this new point of view seems desirable given the high complexity of multiple scattering situations. In practice, it might require considerably more elaborate measurements. 

\bibliographystyle{IEEEtran}
\bibliography{DecompLG_IEEEIT}





\end{document}